\documentstyle[twoside,fleqn,espcrc2]{article}

\newcommand{\AmS}{{\protect\the\textfont2
  A\kern-.1667em\lower.5ex\hbox{M}\kern-.125emS}}

\title{What is the Scale of Supersymmetry Breaking?}

\author{Michael Dine
\address{Physics Department, University
of California, Santa Cruz\\ 
  Santa Cruz, CA    95064}
\thanks{Work supported by the U.S. Department
of Energy.To appear in Proceedings of Fourth International Conference on
Supersymmetry (SUSY96).}}
       
\begin{document}
\begin{abstract}
For a long time, it has been widely assumed that if the underlying
laws of nature are supersymmetric, supersymmetry is broken
at a scale intermediate between the weak scale and the Planck mass.
The construction of realistic models of dynamical supersymmetry
breaking in which supersymmetry is broken at a much lower
scale, as well as a growing appreciation
of the supersymmetric flavor problem has reopened this question.
After reviewing some ideas
for understanding the microscopic origin of
the soft breaking parameters in the context of string theory,
I turn to low energy breaking.  Independent of the details
of the underlying theory,
events with photons and missing energy,
like the CDF  $e^+e^-\gamma \gamma \not{E}_T$ event,
are likely signals of low energy breaking.  I briefly
review the predictions of the simplest model
of low energy supersymmetry breaking
for the soft breaking parameters (MGM), and
then ask what sorts of generalizations are possible.
It turns out that if all couplings are weak,
there are only a limited number of ways
to modify the model.
\end{abstract}

\maketitle

\section{Introduction}

In thinking about supersymmetry phenomenology, there are
(at least) three reasonable approaches to the question:
what is the origin of
supersymmetry breaking?  The first is to ignore the
question, and simply study a supersymmetric
theory with explicit soft breakings\cite{dg}.
This approach,
while pragmatic, can hardly be predictive.  In the case
of the minimal supersymmetric standard model, for
example, there are $106$ parameters.  Typically,
in order to make progress,
one makes a simple ansatz for these, such as degeneracy
of the squark mass matrix at some high energy scale
and proportionality of the soft trilinear couplings to the
fermion Yukawa couplings.  This automatically satisfies
the constraints from various flavor changing processes.

A closely related approach is based on $N=1$ supergravity.
Here one assumes supersymmetry breaking in a hidden
sector, with $F$ and $D$ components
of some chiral and vector fields obtaining vev's
at a scale of order $10^{11}$ GeV\cite{nilles}.
However, these models
still possess 106 parameters, which correspond to terms in the
Kahler potential and gauge coupling functions involving
hidden sector fields.  Typically one assumes
again degeneracy and proportionality
at the highest energy scale.  However,
only if one has a microscopic theory
which explains these parameters does one
have a truly predictive framework, in which one
can evaluate the plausibility of such assumptions.\footnote{
Mass differences in these theories receive ultraviolet
divergent radiative corrections, as is
appropriate for {\it parameters} of a model.}
Of course, the only microscopic theory
of (super) gravity which
we presently know is string theory.  As I will briefly review,
within our current understanding, string theory offers
support both to optimists and pessimists on this issue.
Alternatively, one can ask whether, lacking a detailed
understanding, symmetries might help in solving the supersymmetric
flavor problem and reducing the number of parameters\cite{symm1}
\cite{symm2}\cite{symm3}.
Here, there has been somewhat greater
success, as we have heard at this meeting\cite{symm4}.

The third possibility is that supersymmetry is broken
at low energies,  and that gauge interactions are the
``messengers" of supersymmetry breaking\cite{early}\cite{dns}
\cite{dnns}.
As I will
discuss in more detail, this approach has a number of virtues:

\begin{itemize}
\item  It is highly predictive.  The soft breakings can
all be computed in terms of two or three parameters
\item  The degeneracies required to suppress
flavor-changing neutral currents are automatic.
\item  This framework readily incorporates dynamical
supersymmetry breaking, and thus offers promise of explaining
the hierarchy.
\item  For a broad range of parameters, low energy,
gauge-mediated supersymmetry breaking makes distinctive
and dramatic experimental predictions.
\end{itemize}

Existing models still suffer from certain drawbacks.   Perhaps
the most serious of these is the $\mu$ problem\cite{dns}\cite{dnns}
\cite{pomarol}.
While several
solutions have been offered, none yet seems compelling
(see the talk by Pomarol at this meeting for a discussion
of this problem).

\section{Supersymmetry Breaking in String Theory}

All known classical string vacua possess moduli.  The problem
of supersymmetry breaking in string theory is the problem
of undersanding the lifting of these flat directions.  Inevitably,
whatever potentials are generated -- by instantons, gaugino
condensation, or other mechanisms -- fall to zero in regions
of the moduli space where a weak coupling description
is possible\cite{ds}.
Thus stabilization of the moduli
must occur, if it occurs at all, in regions of strong coupling.

But there are phenomenological
reasons to think that if string theory
does describe nature it sits far from the region of
weak coupling.  In particular, one's naive expectation is that
if $R$ is the compactification radius and $T$ the string tension,
\begin{equation}
\alpha_{GUT}= {\alpha_{string} \over R^6 T^3}
\end{equation}
If we identify $R=M_{GUT}^{-1}$,
 then $\alpha_{string}>10^6$!  The various
known duality transformations take the heterotic
string at such very strong coupling and large radius to moderate
coupling in other theories\cite{dineshirman}.
Witten has pointed out that
in the large radius limit one expects that ``M theory," should
provide a better description of physics than
weak coupling strings\cite{wittenstrong}.
>From the perspective of $M$-theory, the relevant parameters are
the eleven dimensional Planck mass, $M_{11}$,
the radius of the eleventh dimension, $R_{11}$,
and the compactification radius (unification scale), $R$.
These first two quantities can be determined in terms of the
four dimensional Planck mass and the gauge coupling
(interpreted as the unified coupling at the scale $R$).
One finds, taking $R^{-1} \approx 3 \times 10^{16} GeV$,
that 
\begin{equation}
R_{11} M_{11} \approx 8 ~~~~~~~~
RM_{11} \approx 2.
\end{equation}
While it is not necessarily germane to the
issue of universality which we are addressing
here, it is worth pausing to note
that this has the striking implication that all of
the important scales of the theory are of order $M_{GUT}$.
This fact has implications for proton decay,
the strong CP problem, and other questions.

What does string theory  have to say about the question
of universality?  In the weak coupling limit, it has been
known for some time that, if the dilaton dominates
supersymmetry breaking, one gets approximate universality
\cite{ibanezetal}.  It turns out that in the eleven dimensional
limit, one also obtains universality under certain
conditions\cite{banksdine}.
In both cases, taking the weak coupling picture literally,
the degree of universality is at best just barely consistent
with the bounds from flavor changing effects\cite{louisnir}.  However,
it is unlikely that stabilization of the moduli
can occur in a regime where the coupling is weak; corrections
to the Kahler potentials of the various fields -- and,
as a result, the corrections to the soft breaking
terms -- are almost certainly large.
In both the very weak
and very strong coupling limit, then, the assumption
of weak coupling is incompatible
with stabilization of the moduli\cite{banksdine}.
One can summarize the situation
of high scale breaking in string theory
by saying that there are hints
for possible sources of universality, but no clear and
compelling picture really exists.  Flavor symmetries, at
present, seem to provide a more promising solution
of the supersymmetric flavor problem in the context
of high scale breaking.

\section{Low Energy Dynamical Supersymmetry Breaking}

Over the last few years, realistic
models have been constructed in which
supersymmetry is dynamically broken at low
energies\cite{dns,dnns}.
In these models, the messengers of supersymmetry
breaking are ordinary gauge interactions (``Gauge Mediated
Superymmetry Breaking," or GMSB).  In the minimal
model of this kind (MGM), the messengers have the
quantum numbers of a $5$ and $\bar 5$ of $SU(5)$.
They couple to a singlet field, $S$, through a superpotential
\begin{equation}
k_1 S q \bar q + k_2 S l \bar l.
\end{equation}
Due to interactions with some supersymmetry breaking
sector of the theory, the
scalar and $F$-components
of $S$, $\langle S \rangle$ and $\langle F_S \rangle$
are non-vanishing.  Gaugino masses arise at one
loop; scalar masses at two loops.  They are given respectively,
by the expressions:
\begin{equation}
m_{\lambda_i}=c_i\ {\alpha_i\over4\pi}\ \Lambda\ ,
\label{gauginomasses}
\end{equation}

\begin{equation}
\tilde m^2 ={2 \Lambda^2} 
 [~
C_3\left({\alpha_3 \over 4 \pi}\right)^2
+C_2\left({\alpha_2\over 4 \pi}\right)^2 
\label{scalarmasses}
\end{equation}

\noindent
$$
~~~~~~~~~~~~~
+{5 \over 3}{\left(Y\over2\right)^2 \left({\alpha_1\over 4 \pi}\right)^2
}  ],$$

\noindent
where $\Lambda=\langle F_S \rangle /\langle S \rangle$,
$C_3 = 4/3$ for color triplets and zero for singlets,
$C_2= 3/4$ for
weak doublets and zero for singlets, 
and $Y$ is the ordinary hypercharge.\footnote{These formulas
predict a near degeneracy of the bino and the
right handed sleptons.  Important corrections due
to operator renormalization and $D$ terms have
been discussed in \cite{wilczek}.} In this model,
the $A$ terms are high order effects.   The $\mu$
and $B\mu$ parameters can arise through interactions of the
Higgs fields with various singlets\cite{dnns}.

This model is quite simple.  At low energies, the only
new parameters relative to the minimal standard model
are $\Lambda$, $\mu$
and $B \mu$.  Because in the leading approximation
scalar masses are functions only of gauge quantum
numbers,
flavor changing processes are automatically
suppressed.  A negative mass for $H_U$ arises through
top quark loops and gives rise to
$SU(2) \times U(1)$ breaking, in a manner
similar to that in more conventional theories.
Because one does not need to run the
renormalization group equations over too many decades
in energy, one can write the following approximate
expression for the Higgs mass:
\begin{equation}
$$(m_{H_U})^2 = (m_{H_U}^o)^2 - {6 y_t^2 \over 16
\pi^2} \ln(\Lambda^2/\tilde m_t^2)(m_{t}^o)^2
\label{higgsmass}
\end{equation}
Note that while the coupling is not so large,
$(m_{t}^o/m_{H_U}^o)^2\sim 20$.  So the
mass of $H_U$ is negative\cite{ibanezross,acw}.

If we suppose that the lightest charged leptons have
masses of order $100$ GeV, then $\Lambda$ must be greater
than about $30$
TeV. In what follows, we will suppose that $\Lambda^2$
is of order the Goldstino
decay constant, $F$.  This assumption
is the most natural one, but it should
be kept in mind that the scale can be substantially
larger.  Indeed, in many existing models this
scale {\it is} larger by an orders of magnitude or more.
We will see that the phenomenology of the theory
depends sensitively on the value of $F$\cite{ddrt}\cite{swy}.

Phenomenologically, the most distinctive feature of these
models is that the LSP is the gravitino.   The next to lightest
supersymmetric particle (NLSP) is typically a neutralino
or charged, right-handed slepton.  This particle will
decay to its superpartner plus a gravitino, with a
rate governed by low energy theorems.  Over a large
region of the parameter space, the NLSP is neutral,
with a significant bino component.  The lifetime
is then given by\cite{fayet}:
\begin{equation}
\Gamma( \tilde{B} \rightarrow G + \gamma) = 
{ \cos^2 \theta_W ~ m_{\tilde{B}}^5 \over 16 \pi F^2}
\label{decayrate}
\end{equation}
This translates to a decay length
\begin{equation}
c \tau \simeq 1.3 \times 10^{-2} 
\left(100~{\rm GeV}  \over m_{\tilde{B}} \right)^5
\left(  \sqrt{F} \over  100~{\rm TeV}\right)^4 ~{\rm cm}
\label{decaylength}
\end{equation}
In other words, for $F$ up to about $1000$ TeV,
the decay occurs in the detector.  For a significant
range of $F$, there is the possibility of measuring
displaced vertices\cite{ddrt}\cite{swy}.
Note that a formula like eqn. 7 will hold in {\it any}
model of low energy supersymmetry breaking,
so long as the NLSP has a significant gaugino
component.

We have already heard a good deal at this meeting about
the experimental signatures for such processes.
Assume, for the moment, that the bino is the NLSP.   Then
in $e^+ e^-$ annihilation, bino pairs can be produced
(with selectron exchange) leading to final states with
two photons and missing energy.  This process will be
used during the upcoming LEP runs to set limits on the
bino mass.  Similarly, in $p \bar p$ collisions, one
can produce various new particles, yielding final
states with photons and missing energy.  As has
been discussed at this meeting in the talks
by Conway, Kane and Thomas, there is a candidate event
of this type in the CDF data sample.  Thomas, in particular,
has discussed in some detail the interpretation of this
event within the framework of low energy supersymmetry
breaking\cite{dtw}.

So far, we have discussed a particular model for the messenger
sector.  This model is highly predictive.
But while events with photon pairs and missing energy seem
to be a generic signature of low energy supersymmetry
breaking, it is less clear that the detailed predictions for the
spectrum are generic.  It turns out, however, that if the
underlying theory is weakly coupled, there are only
a limited set of possibilities for the messenger sector.
One can reason as follows:
\begin{itemize}
\item
At tree level, there are some rules for the spectrum which
insure, for example, that there is a squark with mass
lighter than the $u$-quark\cite{dg}.  As a result,
SUSY breaking must be fed through loops.
\item
Some of the messengers must carry color, or the gluino is too light.
\item
Because the messenger fields are more massive
than the weak scale by powers of $\alpha$,
they must fall in vector representations of $SU(3)\times SU(2)
\times U(1)$.
\item
Unification requires that the messengers fit into complete $SU(5)$ 
representations.
\item
Requiring that no couplings blow up before
the unification scale constrains the number of messengers.
One can have at most four $5+ \bar 5$'s,
or one $10 + \bar 10$.
\item
The messengers must couple to singlets, with non-vanishing
scalar and $F$-components.
\end{itemize}

In other words, the model we have written is the simplest of a
narrow class of models.  The modifications we have described here
spoil the mass formula in detail, but not universality.  There
is one other possible modification of the theory which
we have not yet considered.\footnote{The remarks
which follow have been developed in discussions with
Savas Dimopoulos, Yossi Nir,
Yuri Shirman and Scott Thomas.}  In the simple model, we insisited
that there was no mixing of messengers with ordinary
matter fields.  This can be enforced by discrete symmetries,
but it is not necessarily true.  For
example, one can contemplate couplings such as
\begin{equation}
H_D \bar d y_d  Q + H_D \bar q Y_d Q.
\label{mixingterms}
\end{equation}
Such couplings do not alter the KM structure.  But
they do imply new contributions to squark
and slepton masses.  A simple one loop computation
yields a negative mass shift:
\begin{equation}
\delta m^2 = -{Y^2 \over 16 \pi^2}{(F^{\dagger} F)^2 \over 6 M^6}.
\label{nonuniversalshift}
\end{equation}
These contributions are certainly non-universal.  However,
our experience with ordinary quarks and leptons suggests
that at most one or two states would be significantly
displaced from the positions implied by eqn. 5.

Finally, what if the supersymmetry breaking sector
is strongly coupled?  In this case, the rules are
far less clear.  For example, Seiberg\cite{seibergduality}
has taught us that
theories in the infrared may look quite different
than in the ultraviolet, so perhaps we should drop our
insistence on asymptotic freedom.  Also, in strongly
interacting theories, it is not so clear how the mass
formulas may look.  For example, one might expect
an equation like 4 to hold, but perhaps
not with so many factors of $4 \pi$.  In such theories,
scalars might be lighter than gauginos.\footnote{
I thank Scott Thomas for raising this issue, and for discussions.}

We have seen in recent years that low energy dynamical
supersymmetry breaking is quite plausible.
While we do not yet have a model which, like
the original Weinberg-Salam model, is compelling
in its simplicity, progress in the understanding of
supersymmetric dynamics raise hopes that this will
be achieved.  Given that photons plus missing
energy are a generic consequence of this
framework, confirmation of the CDF
observation would provide a strong impetus
to find this model.

\end{document}